\documentclass[aps,prd,reprint,superscriptaddress,nofootinbib,floatfix]{revtex4-1}

\usepackage{amsfonts,amssymb,amsmath}
\usepackage{epsfig}
\usepackage{bm}
\usepackage{slashed}
\usepackage{latexsym}
\usepackage{natbib}
\usepackage{url}
\usepackage{dcolumn}
\usepackage{color}
\usepackage{graphicx}
\usepackage{psfrag}
\usepackage{ulem}
\usepackage{hyperref}
\hypersetup{colorlinks   = true,
            urlcolor     = blue,
            citecolor    = blue,
            linkcolor    = blue,
            menucolor    = blue,
            anchorcolor  = blue,
            filecolor    = blue
}
\everymath{\displaystyle}

\begin{document}

\title{Transient axion streams from disrupted miniclusters}

\author{Luca Visinelli}
\email{lvisinelli@unisa.it}
\affiliation{Dipartimento di Fisica ``E.R.\ Caianiello'', Universit\`a degli Studi di Salerno,\\ Via Giovanni Paolo II, 132 - 84084 Fisciano (SA), Italy}
\affiliation{Istituto Nazionale di Fisica Nucleare - Gruppo Collegato di Salerno - Sezione di Napoli,\\ Via Giovanni Paolo II, 132 - 84084 Fisciano (SA), Italy}

\author{Momchil Naydenov}
\email{mnaydenov@phys.uni-sofia.bg}
\affiliation{Faculty of Physics, Sofia University, 5 J.\ Bourchier Blvd, 1164 Sofia, Bulgaria}
\affiliation{Istituto Nazionale di Fisica Nucleare, Laboratori Nazionali di Frascati, C.P. 13, 00044 Frascati, Italy}

\begin{abstract}
We investigate the formation and evolution of axion streams generated by the tidal disruption of axion miniclusters through stellar encounters in the Milky Way halo. Combining a large-scale Monte Carlo treatment of repeated stellar flybys with a tracer reconstruction of the stripped debris, we follow the phase-space evolution of the streams across a broad range of galactocentric radii and assess their contribution to the local dark matter distribution. We find that the kinetic energy of the stripped debris typically exceeds its residual self-gravitational binding energy at formation, so that the subsequent evolution is dominated by anisotropic free expansion and orbital shear. As a result, stream densities can decrease by factors as large as $\sim10^{-9}$ over Galactic timescales, strongly suppressing the steady-state abundance of dense streams near the Solar circle. At the Solar radius, only a small fraction of realizations yields a nonzero encounter probability over a 10-year exposure, implying that observable streams are dominated by rare recent and nearby disruption events rather than by a persistent population of long-lived overdense substructure. Despite this rapid dilution, the streams remain dynamically cold and produce detector-frame linewidths many orders of magnitude narrower than the cavity bandwidths of current haloscope experiments. For representative haloscope configurations, we find characteristic stream linewidths in the range $\Delta\nu_{\rm stream}\sim10^{-7}$--$10^{1}\,{\rm Hz}$, while the corresponding Doppler drift remains well below the cavity response width.
\end{abstract}

\date{\today}
\maketitle

\section{Introduction}

The QCD axion~\cite{Weinberg:1977ma,Wilczek:1977pj} is one of the most compelling candidates for the dark matter (DM) of the Universe, arising naturally from the Peccei--Quinn (PQ) solution to the strong CP problem~\cite{Peccei:1977hh,Peccei:1977ur}; see Refs.~\cite{Marsh:2015xka,DiLuzio:2020wdo,Arza:2026rsl} for reviews. In post--inflationary PQ symmetry-breaking scenarios, spatial variations of the axion field generate $\mathcal{O}(1)$ density fluctuations at the onset of oscillations. These fluctuations can become nonlinear and gravitationally collapse into compact bound structures known as axion miniclusters (AMCs)~\cite{Hogan:1988mp,Kolb:1993hw,Kolb:1993zz}. Subsequent studies have explored the formation, survival, and phenomenology of AMCs, including their disruption in Galactic environments and their contribution to the Milky Way DM halo~\cite{Tkachev:2014dpa, Tinyakov:2015cgg}.

Over the past decade, significant progress has been made in understanding the formation and internal structure of AMCs. Simulations of axion string and domain wall networks have refined predictions for the axion relic abundance and the small-scale density fluctuations relevant for AMC formation~\cite{Hiramatsu:2012sc,Hiramatsu:2012gg, Fleury:2015aca, Klaer:2017ond,Gorghetto:2018myk, Buschmann:2019icd,Gorghetto:2020qws, Buschmann:2021sdq, Pierobon:2023ozb, Benabou:2024msj, Saikawa:2024bta}, while dedicated studies of nonlinear structure formation have characterized the AMC mass function and internal density profiles~\cite{Fairbairn:2017sil,Enander:2017ogx,Vaquero:2018tib}. These results indicate that AMCs span a broad mass range, typically $M_{\rm AMC}\sim10^{-16}$--$10^{-10}\,M_\odot$, with densities far above the cosmological mean. Their internal structure is often approximated either by Navarro--Frenk--White profiles or by steep power-law profiles when the inner scale radius is unresolved~\cite{Vaquero:2018tib,Kavanagh:2020gcy,Ellis:2022grh}. Although the detailed properties of AMCs depend on the cosmological history and axion model parameters~\cite{Blinov:2019rhb, Visinelli:2020hat, Choi:2022btl, Hardy:2026mkp}, their existence is a robust prediction of post--inflationary axion cosmology. More generally, axion direct-detection experiments are sensitive to fine-grained features of the Galactic halo, including transient overdensities and departures from a smooth Maxwellian velocity distribution~\cite{Foster:2017hbq}.

Once formed, AMCs evolve within the Galactic environment under the combined influence of the smooth Galactic tide and impulsive stellar encounters, which can induce heating, mass loss, and eventual disruption~\cite{Green:2006hh,Kavanagh:2020gcy,Edwards:2020afl, Walters:2024vaw}. Monte Carlo calculations specific to AMCs have shown that their survival probability depends sensitively on the internal density profile and orbital history, with compact systems surviving preferentially while diffuse objects are efficiently stripped~\cite{Kavanagh:2020gcy}. Idealized $N$-body simulations of related CDM microhalos and minihalos have provided complementary calibrations of their response to stellar heating and Galactic tides~\cite{Schneider:2010jr, Delos:2019tsl, Shen:2022ltx}. Although the progenitor structures considered in these studies are not identical to the PL AMCs adopted here, they provide the closest available dynamical comparisons. Additional effects may also be important: wave dynamics can modify the response of axion structures to tidal perturbations~\cite{Dandoy:2022prp}, while calculations of successive stellar encounters show that cumulative energy injection and finite relaxation between impulses can enhance mass loss~\cite{DSouza:2024flu, DSouza:2024uud}.

Mass loss naturally produces extended tidal debris. Direct simulations of disrupted CDM microhalos show that stellar heating and the Galactic tide can generate cold, filamentary debris structures~\cite{Schneider:2010jr}. In the AMC context, the stripped component has similarly been modeled as parsec-scale streams with potentially observable local-density enhancements~\cite{Tinyakov:2015cgg, OHare:2023rtm}. More generally, the evolution of collisionless tidal debris in Galactic potentials is governed by orbital shear and phase mixing and is naturally described in action--angle space~\cite{Helmi:1999ks, Sanders:2013sda, Bovy:2014yba}. This broader stream literature shows that the dilution rate depends on the dimensionality of the frequency-space distribution and on the regular or chaotic character of the underlying orbit~\cite{Price-Whelan:2015uea}. Consequently, establishing that the debris is no longer self-bound does not by itself determine its subsequent density law or coherence time.

Such streams are especially relevant for axion haloscopes~\cite{Sikivie:1983ip, Sikivie:1985yu}, because a dynamically cold component produces a highly non-Maxwellian velocity distribution and a narrow spectral feature~\cite{OHare:2017yze, OHare:2023rtm}. Recent studies have emphasized that fine-grained streams may require spectral resolutions substantially finer than the cavity bandwidth and an explicit treatment of time-dependent Doppler shifts~\cite{ADMX:2023ctd,Yi:2025ktq, OHare:2025jpr}. Related searches and phenomenological studies further illustrate the potential sensitivity of axion experiments to cold substructure~\cite{Bhura:2026bpy, Maroudas:2026ejc}.

Despite this progress, the connection between the event-by-event stripping history of an AMC and the subsequent phase-space evolution of its debris remains comparatively unexplored. Most previous AMC studies have concentrated on the survival and mass loss of the bound remnant, whereas detector forecasts generally begin from an assumed stream geometry or velocity distribution. It therefore remains unclear whether the stripped component can persist as a coherent overdensity over astrophysical timescales or whether its inherited velocity dispersion, encounter-induced perturbations, and Galactic orbital shear rapidly disperse it.

In this work, we revisit the formation and evolution of axion streams generated by stellar encounters with AMCs, focusing on their dynamical stability. We adopt the Monte Carlo framework for stellar disruption developed in Ref.~\cite{Kavanagh:2020gcy}, specialized here to an axion mass $m_a=50\,\mu{\rm eV}$ and power-law (PL) AMC density profiles. For each realization, we track the stripped mass and reconstruct the kinematic properties of the debris, including both longitudinal and transverse velocity dispersions. We then model the subsequent evolution of the stream as a collisionless system to determine whether the resulting structures can survive as persistent overdensities within the Galactic halo. Our analysis indicates that axion streams produced by tidal stripping evolve predominantly through collisionless expansion and phase mixing. For typical stripped mass fractions and stream geometries, the velocity dispersion of the debris exceeds the escape velocity associated with its self-gravity. As a result, the streams undergo anisotropic expansion, with longitudinal stretching driven by tidal kicks and transverse broadening inherited from the progenitor. Within the adopted packet initialization and over the simulated interval, the reconstructed coarse-grained density approaches the full-rank ballistic benchmark $\rho_{\rm stream}\propto t^{-3}$. This rapid dilution suggests that the modeled structures are transient, although their multi-orbit asymptotic evolution requires a more complete treatment in action--angle space or a dedicated self-consistent simulation.

These findings have important implications for the phenomenology of axion DM. In contrast to scenarios in which long-lived streams dominate the local phase-space distribution, we find that tidal debris is typically short-lived and subdominant. Observable streams may still arise from recent and nearby disruption events, although such occurrences are expected to be rare. Our results place the detectability of axion streams within a more constrained dynamical framework, directly linking their observational prospects to the disruption history of miniclusters in the Milky Way. The original code used for the tidal stripping of AMCs was developed in Ref.~\cite{Kavanagh:2020gcy} and is publicly available at \href{https://github.com/bradkav/axion-miniclusters/}{github.com/bradkav/axion-miniclusters}. The extensions and analysis tools developed for the present work, including the stream-evolution and tracer reconstruction modules, are publicly available at \href{https://github.com/lucavisinelli/AxionStreams/}{github.com/lucavisinelli/AxionStreams}.

\section{Physics setup}

We summarize the physical framework describing the formation and tidal disruption of AMCs in a galactic environment, following the Monte Carlo approach developed in Ref.~\cite{Kavanagh:2020gcy}. In post--inflation PQ scenarios, $\mathcal{O}(1)$ density fluctuations at the onset of axion oscillations lead to the formation of gravitationally bound structures shortly after matter--radiation equality~\cite{Hogan:1988mp,Kolb:1993hw}. These AMCs span a broad mass range and are characterized by internal densities determined by the primordial overdensity $\delta$~\cite{Kolb:1993hw},
\begin{equation}
    \rho_{\rm AMC} = 140\,(1+\delta)\,\delta^3\,\rho_{\rm eq}\,,
\end{equation}
where $\rho_{\rm eq}$ is the cosmological matter density at equality. For a given mass and density, the characteristic radius and internal velocity dispersion are
\begin{equation}
    R_{\rm AMC} = \left(\frac{3 M_{\rm AMC}}{4\pi \rho_{\rm AMC}}\right)^{1/3}, 
    \qquad
    \sigma_{\rm AMC} \sim \left(\frac{G M_{\rm AMC}}{R_{\rm AMC}}\right)^{1/2}.
\end{equation}
Alongside AMCs, denser gravitationally bound objects known as axion stars may also form, characterized by different equilibrium configurations and density profiles~\cite{Barranco:2010ib, Chavanis:2011zm, Visinelli:2017ooc, Gorghetto:2024vnp}.

After their formation, AMCs evolve within the Galactic halo under the combined influence of the smooth gravitational potential and discrete encounters with stars. The stellar population is described by a spatially varying number density $n_\star({\bf r})$ and velocity dispersion $\sigma_\star({\bf r})$, which determine both the rate and strength of gravitational flybys. The characteristic relative velocity is
\begin{equation}
    v_{\rm rel} \sim \sqrt{2}\,\sigma_\star.
\end{equation}
When the duration of an encounter is short compared to the internal dynamical time of the minicluster, the interaction can be treated within the impulse approximation. The corresponding energy injection is
\begin{equation}
    \label{eq:DeltaE}
    \Delta E \sim \frac{G^2 M_\star^2}{b^4 v_{\rm rel}^2}\,\langle r^2 \rangle \, M_{\rm AMC}'\,,
\end{equation}
where $M_\star$ is the stellar mass, $b$ is the impact parameter, and $\langle r^2\rangle$ is the mean-square radius of the minicluster. Here, $M_{\rm AMC}'$ is the instantaneous mass of the AMC. Repeated encounters lead to cumulative heating, tidal stripping, and progressive mass loss from the outer regions of the bound object.

Each realization is initialized by assigning an AMC with initial mass $M_{\rm ini}$ and density $\rho_{\rm ini}$ drawn from a prescribed mass function. The corresponding radius $R_{\rm ini}$ and velocity scale $\sigma_{\rm AMC}$ follow from the relations above. The minicluster is then placed on a bound Galactic orbit specified by its semi-major axis, eccentricity, and orbital phase, which determine the local stellar environment encountered along the trajectory.

The cumulative effect of stellar flybys is modeled as a sequence of discrete perturbations with encounter rate
\begin{equation}
    \frac{{\rm d}N}{{\rm d}t} = n_\star({\bf r})\, \pi b_{\max}^2\, v_{\rm rel}(t)\,,
\end{equation}
where ${\bf r}={\bf r}(t)$ and $b_{\max}$ is an effective maximum impact parameter. Individual encounters are sampled from the corresponding local distributions of impact parameters and relative velocities. For each event, the injected energy is computed using Eq.~\eqref{eq:DeltaE}, and the subsequent dynamical response is implemented through a semi-analytic stripping operator that updates the bound mass and radius of the AMC after energy injection and structural readjustment. Iterating this procedure yields the full stripping history of each realization. Energy-injection prescriptions of this general type have been tested against idealized $N$-body stellar encounters for compact microhalos and minihalos~\cite{Delos:2019tsl, Shen:2022ltx}. Their quantitative application to the PL AMC population adopted here remains a modeling assumption.

Relative to the original implementation of Ref.~\cite{Kavanagh:2020gcy}, the framework adopted here retains detailed information about the stripped material itself rather than focusing only on the surviving bound remnant. Each stripping event is recorded individually together with its associated orbital and kinematic properties, including the stripping time, local phase-space configuration, injected energy, stripped mass, and characteristic velocity scales inherited from the tidal encounter. This allows the debris to be reconstructed as a sequence of localized packets in phase space rather than as a single effective stream component. The framework also reconstructs the longitudinal and transverse velocity dispersions associated with each stripping event, which provide the initial conditions for the subsequent stream evolution developed below. The main output of the Monte Carlo stage is the event history of each realization, consisting of a sequence of stripping events indexed by $i$ and characterized by a stripping time $t_i$, stripped mass element $\Delta M_i$, orbital position ${\bf x}_i$, velocity ${\bf v}_i$, and the corresponding effective velocity dispersions. These stripping histories provide the initial conditions for the stream reconstruction and phase-space evolution discussed in the following section.

\section{Stream reconstruction}

To follow the evolution of the stripped debris, we extend the Monte Carlo framework of Ref.~\cite{Kavanagh:2020gcy} by retaining the full stripping history of each AMC realization. Rather than characterizing disruption only through the surviving bound fraction, we reconstruct the stripped component as a collisionless tracer ensemble evolving in the Galactic potential. This allows us to assess if the resulting axion streams can remain coherent and self-gravitating over Galactic timescales, or instead disperse through phase mixing and orbital shear.

The numerical framework consists of three stages. First, the Monte Carlo evolution follows AMCs undergoing repeated stellar encounters and records the stripping history of each realization. Second, the stripped material is reconstructed as a collection of localized phase-space packets. Third, tracer particles associated with these packets are evolved in a smooth Galactic potential and used to infer the stream morphology, density evolution, visibility time, and effective abundance. The tracer ensemble should be interpreted as an effective coarse-grained representation of the stripped debris rather than a fully self-consistent $N$-body realization. Unlike the original Monte Carlo framework of Ref.~\cite{Kavanagh:2020gcy}, which follows only the evolution of the bound AMC remnant, our implementation stores the full stripping history of each realization. Each stripping event with nonzero mass loss is characterized by the stripping time $t_i$, stripped mass $\Delta M_i$, injected energy $\Delta E_i$, and the local orbital configuration of the parent AMC. Since the Monte Carlo calculation does not explicitly evolve the stripped particles, we reconstruct their initial phase-space coordinates from the local orbital properties of the AMC at the stripping event. Each event is described by a localized packet with phase-space coordinates $({\bf x}_i,{\bf v}_i)$ corresponding to the effective bulk motion of the debris immediately after stripping. The packet carries a total mass $\Delta M_i$ and seeds the subsequent tracer reconstruction.

The internal phase-space structure of the packet is modeled through an anisotropic Gaussian velocity distribution aligned with the local stream direction. The longitudinal velocity dispersion is estimated from the specific energy injected during the encounter,
\begin{equation}
    \sigma_{l,i}^2 \simeq \frac{2\,\Delta E_i}{M_i}\,,
\end{equation}
where $M_i$ is the bound AMC mass immediately before stripping. The transverse velocity scale is identified with the characteristic internal velocity dispersion of the progenitor AMC,
\begin{equation}
    \sigma_{t,i} \simeq \left(\frac{G M_i}{R_i}\right)^{1/2}\,,
\end{equation}
where $R_i$ is the corresponding AMC radius. Physically, $\sigma_{l,i}$ parametrizes the velocity spread induced by the tidal encounter, while $\sigma_{t,i}$ is the initial thickness of the stream inherited from the internal dynamics of the parent AMC.

For computational simplicity, the packet centroid is reconstructed using a planar approximation to the local AMC orbit. The sampled orbital radius $r_i$, orbital phase $\phi_i$, and AMC speed $v_i$ determine the initial phase-space coordinates of the packet. A local orthonormal basis is then constructed from the packet velocity, defining one longitudinal direction parallel to the bulk motion and two transverse directions orthogonal to it. Tracer particles are sampled around the packet centroid using the dispersions $(\sigma_{l,i},\sigma_{t,i})$ along these directions. The tracers are subsequently evolved in an external smooth Galactic mean field $\Phi({\bf x})$, according to
\begin{equation}
    \ddot{\bf x} = -\nabla\Phi({\bf x})\,,
\end{equation}
using a symplectic leapfrog integrator. The tracer particles are evolved in a static logarithmic Galactic potential,
\begin{equation}
    \Phi(r)=\frac{1}{2}v_c^2 \ln\!\left(r^2+r_{\rm soft}^2\right)\,,
\end{equation}
with circular velocity $v_c=220\,{\rm km\,s^{-1}}$ and softening radius $r_{\rm soft}=0.05\,{\rm pc}$, so that an approximately flat rotation curve is produced for $r\gg r_{\rm soft}$. The purpose of this evolution is to capture the effects of orbital shear and phase mixing on the stripped debris. The self-gravity of both the debris and the surviving AMC remnant is neglected. This approximation is justified a posteriori by the fact that the kinetic energy of the reconstructed streams dominates over residual self-gravity, see Eq.~\eqref{eq:alpha} below, so that their subsequent evolution is governed primarily by collisionless expansion and phase mixing.

The tracer evolution naturally produces elongated filamentary structures through orbital shear and phase mixing. To quantify the dilution of the streams, we reconstruct the density using two complementary estimators. The first is a coarse-grained estimator based on the covariance matrix of the tracer distribution,
\begin{equation}
    C_{ij}(t) = \left\langle\bigl(x_i-\langle x_i\rangle\bigr)\bigl(x_j-\langle x_j\rangle\bigr)\right\rangle\,,
\end{equation}
from which we define the effective volume
\begin{equation}
    V_{\rm eff}(t) \equiv \frac{4\pi}{3}\sqrt{\lambda_1\lambda_2\lambda_3}\,,
\end{equation}
where $\lambda_i$ are the eigenvalues of $C_{ij}$, ordered such that $\lambda_1\ge \lambda_2\ge \lambda_3$. The corresponding coarse-grained density estimator is
\begin{equation}
    \label{eq:rhocg}
    \rho_{\rm cg}(t) \sim \frac{M_{\rm stream}}{V_{\rm eff}(t)}\,.
\end{equation}
We also construct a nearest-neighbor density estimator, $\rho_{\rm local}$, based on the distance to the $k$th nearest tracer, which is more sensitive to localized overdensities within the filamentary stream structure. The final stream associated with a given AMC realization is obtained by superposing the tracer populations generated by all stripping events occurred during the disruption.

The density evolution is characterized through the logarithmic slope
\begin{equation}
    \label{eq:gamma}
    \gamma(t) \equiv \frac{{\rm d}\log \rho}{{\rm d}\log t}\,,
\end{equation}
where $\rho$ denotes either $\rho_{\rm cg}$ or $\rho_{\rm local}$. We compare the reconstructed evolution with the ballistic estimate
\begin{equation}
    l_s(t)\sim \sigma_l t, \qquad R_s(t)\sim \sigma_t t\,,
\end{equation}
where $\sigma_l$ and $\sigma_t$ are characteristic longitudinal and transverse velocity scales representative of the tracer ensemble. This implies
\begin{equation}
    \label{eq:rho_ballistic}
    \rho_{\rm stream}(t) \sim \frac{M_{\rm stream}}{\pi R_s^2(t)\,l_s(t)}\propto t^{-3}\,.
\end{equation}
This scaling is not imposed in the simulations and serves only as a benchmark for interpreting the reconstructed stream evolution. The ballistic estimate in Eq.~\eqref{eq:rho_ballistic} assumes that the debris expands independently in three configuration space directions. Over longer intervals in a bound Galactic potential,
stream evolution is more naturally described in action--angle variables, for which the angular separation evolves as $\Delta\bm{\theta}(t)=\Delta\bm{\theta}_0 + \Delta\bm{\Omega}\,t$~\cite{Helmi:1999ks, Sanders:2013sda, Bovy:2014yba}. The resulting density law is controlled by the effective dimensionality and eigenvalue hierarchy of the frequency-space distribution. Thin tidal streams are commonly close to one-dimensional in frequency space, so that their mean density may initially decrease as $t^{-1}$ and later transition to $t^{-2}$ or $t^{-3}$ as additional frequency directions become important. Chaotic frequency evolution can instead produce faster, approximately exponential dispersal~\cite{Price-Whelan:2015uea}. Accordingly, the $t^{-3}$ behavior found here should be interpreted as the full-rank ballistic benchmark realized by the adopted packet model over the simulated interval, rather than as a universal asymptotic law.

To assess whether the stripped component remains self-gravitating, we compare the characteristic specific kinetic energy,
\begin{equation}
    E_{\rm kin} \sim \frac12\left(\sigma_l^2+2\sigma_t^2\right)\,,
\end{equation}
with the specific gravitational binding energy,
\begin{equation}
    E_{\rm bind} \sim \frac{G M_{\rm stream}}{R_s(t)}\,.
\end{equation}
We define the virial parameter
\begin{equation}
    \label{eq:alpha}
    \alpha \equiv \frac{2E_{\rm kin}}{E_{\rm bind}}\,,
\end{equation}
which quantifies the relative importance of kinetic support and residual self-gravity. For the representative stripping realizations reconstructed here we typically find $\alpha\gg1$ already at early times, indicating that the subsequent evolution is dominated by collisionless expansion and phase mixing rather than gravitational collapse. Residual self-gravity may nevertheless remain relevant for unusually dense or marginally bound streams in the high-density tail.

Finally, we define a visibility time $t_{\rm vis}$ as the time at which the reconstructed stream density falls below a threshold $10\,\rho_{\rm DM,loc}$, where $\rho_{\rm DM,loc}$ is the local
Galactic DM density. The corresponding effective dilution rate is
\begin{equation}
    \label{eq:Gammadiffuse}
    \Gamma_{\rm diffuse} = t_{\rm vis}^{-1}\,,
\end{equation}
which provides an estimate of the steady-state stream abundance,
\begin{equation}
    \label{eq:nss}
    n_{\rm stream} \sim n_{\rm AMC}\frac{t_{\rm vis}}{T_{\rm age}} \sim \frac{n_{\rm AMC}}{T_{\rm age}\Gamma_{\rm diffuse}}\,.
\end{equation}
This relation connects the production of tidal debris to its subsequent dilution through phase mixing and orbital shear.

\section{Results}

We now present the results obtained from the Monte Carlo evolution of tidally disrupted AMCs and from the tracer reconstruction of the resulting debris streams. All radial statistics shown below are derived from ensembles with $N_{\rm AMC}=10^6$ realizations at each orbital radius. The sample is sufficiently large to probe the rare-event tails associated with strong stellar encounters and highly disrupted systems. For each orbital configuration, the minicluster population is evolved under repeated stellar encounters while the stripped component is reconstructed through tracer particles. This procedure allows us to follow directly the morphology and density evolution of the debris without imposing an a priori analytic stream model. The orbital evolution is computed within the Milky Way model adopted in Ref.~\cite{Kavanagh:2020gcy}, which includes the dominant stellar populations responsible for tidal perturbations in the Galactic bulge and disk. The simulations probe orbital radii spanning the inner Galaxy, Solar neighborhood, and outer halo, with the semi-major axis ranging within $a = 0.3\text{--}10\,{\rm kpc}$.

\begin{figure}[htb!]
    \centering
    \includegraphics[width=\linewidth]{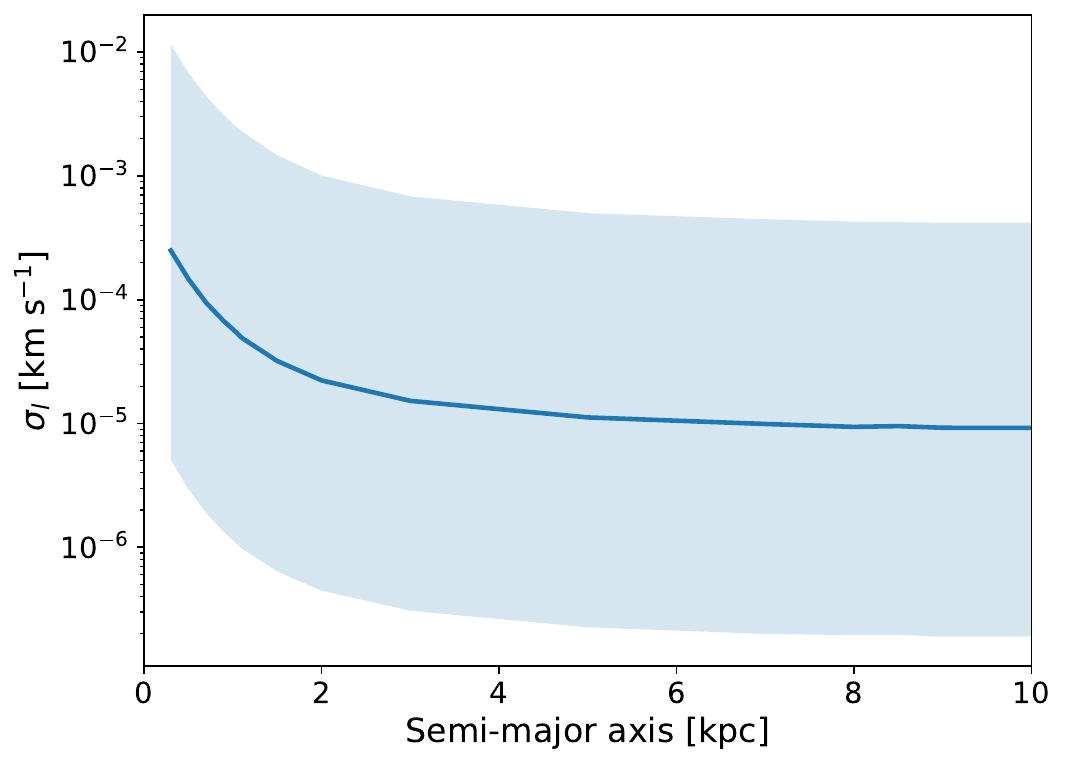}
    \caption{Median longitudinal velocity dispersion $\sigma_l$ of the stripped debris as a function of orbital semi-major axis $a$. The shaded band corresponds to the $1\sigma$ spread among Monte Carlo realizations.}
    \label{fig:sigma_l_vs_radius}
\end{figure}

We begin by considering the longitudinal velocity dispersion $\sigma_l$ induced by tidal stripping and orbital-energy injection. Figure~\ref{fig:sigma_l_vs_radius} shows the median value and statistical spread as functions of galactocentric semi-major axis. The longitudinal velocity dispersion decreases systematically with galactocentric radius. In the inner Galaxy, repeated stellar encounters generate characteristic values $\sigma_l \sim 10^{-4}\,{\rm km\,s^{-1}}$ while, beyond $a\gtrsim5\,{\rm kpc}$, the dispersion asymptotes to $\sigma_l \sim 10^{-5}\,{\rm km\,s^{-1}}$. The broad statistical spread reflects the stochastic nature of the encounter history, particularly for orbits crossing the dense stellar environment of the inner Galaxy.

\begin{figure}[htb!]
    \centering
    \includegraphics[width=\linewidth]{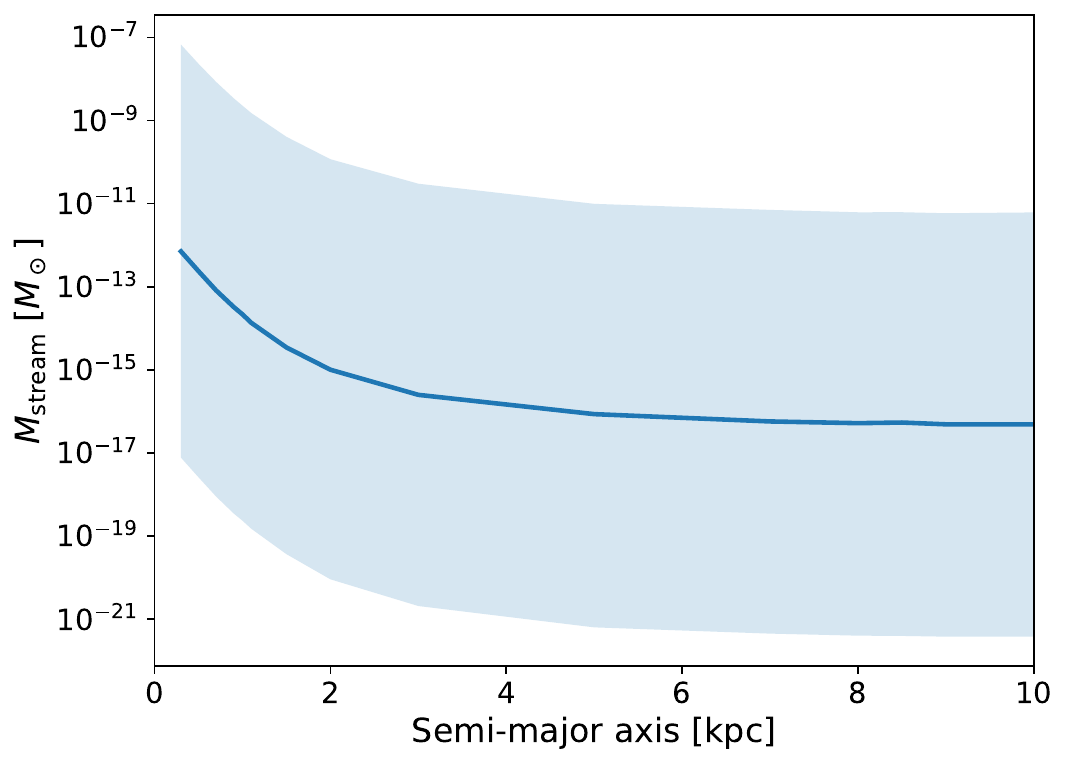}
    \caption{Median stream mass $M_{\rm stream}$ as a function of orbital semi-major axis. The shaded band corresponds to the $1\sigma$ spread among Monte Carlo realizations.}
    \label{fig:Mstream_vs_radius}
\end{figure}

The total stripped mass transferred into the stream,
\begin{equation}
    M_{\rm stream}=M_{\rm ini}-M_{\rm fin},
\end{equation}
also shows a pronounced radial dependence. As shown in Fig.~\ref{fig:Mstream_vs_radius}, the characteristic stripped mass decreases by nearly four orders of magnitude between the inner and outer halo. Typical stream masses are $M_{\rm stream}\sim10^{-13}\,M_\odot$ at sub-kpc radii, while for $a\gtrsim5\,{\rm kpc}$ they decrease to $M_{\rm stream}\sim10^{-16}\text{--}10^{-17}\,M_\odot$. Both $M_{\rm stream}$ and $\sigma_l$ are controlled by the cumulative tidal energy injected through stellar encounters, as indicated by the similarity between their radial profiles. Orbital shear and repeated weak perturbations may broaden the velocity distribution without significantly increasing the stripped mass fraction.

\begin{figure}[htb!]
    \centering
    \includegraphics[width=\linewidth]{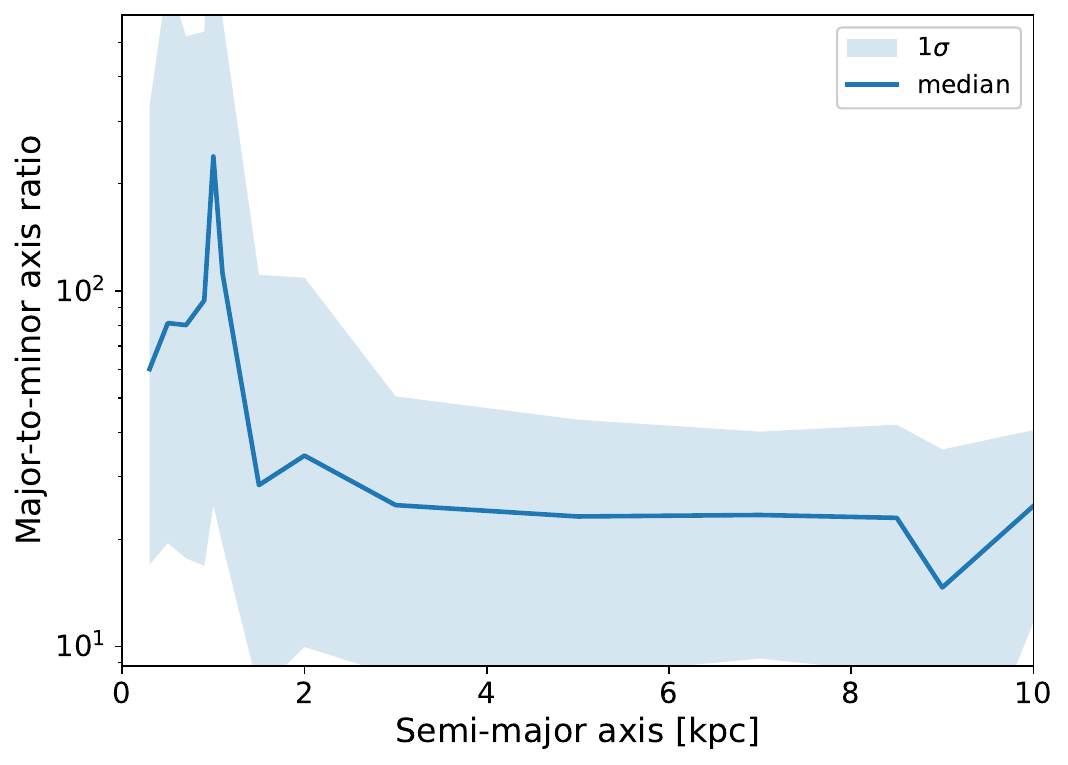}
    \caption{Final major-to-minor axis ratio of the reconstructed stream distribution as a function of semi-major axis.}
    \label{fig:axis_ratio_vs_radius}
\end{figure}

The tracer reconstruction reveals that the stripped debris evolves into highly anisotropic filamentary structures. We characterize the morphology through the ratio between the principal axes of the tracer covariance matrix, shown in Fig.~\ref{fig:axis_ratio_vs_radius}. Streams formed in the inner Galaxy become extremely elongated, with axis ratios reaching $\mathcal{O}(10^2)$ because of strong tidal stretching and orbital shear. Outside the central kiloparsec the anisotropy decreases steadily with galactocentric radius, indicating that streams formed in the outer halo are less filamentary.

\begin{figure}[htb!]
    \centering
    \includegraphics[width=\linewidth]{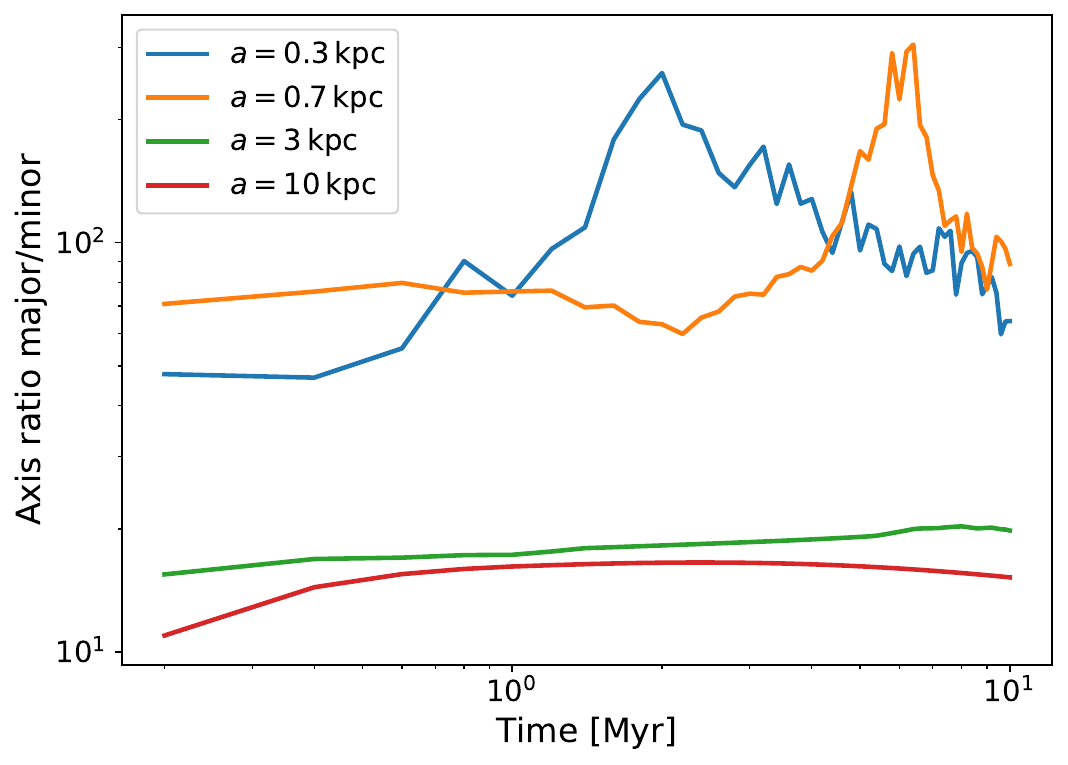}
    \caption{Time evolution of the major-to-minor axis ratio for representative streams at $a=0.3\,{\rm kpc}$, $a=0.7\,{\rm kpc}$, $a=3\,{\rm kpc}$, and $a=10\,{\rm kpc}$.}
    \label{fig:axis_ratio_evolution_compare}
\end{figure}

Figure~\ref{fig:axis_ratio_evolution_compare} shows the time dependence of the major-to-minor axis ratio for representative streams at $a=0.3\,{\rm kpc}$, $a=0.7\,{\rm kpc}$, $a=3\,{\rm kpc}$, and $a=10\,{\rm kpc}$. Streams formed in the inner Galaxy rapidly develop highly elongated morphologies, reaching axis ratios ${\cal O}(10^2)$ within a few Myr. At sub-kpc radii the evolution is irregular because of repeated perturbations and orbital shear, while at larger radii it approaches a smoother ballistic regime.

\begin{figure}[htb!]
    \centering
    \includegraphics[width=\linewidth]{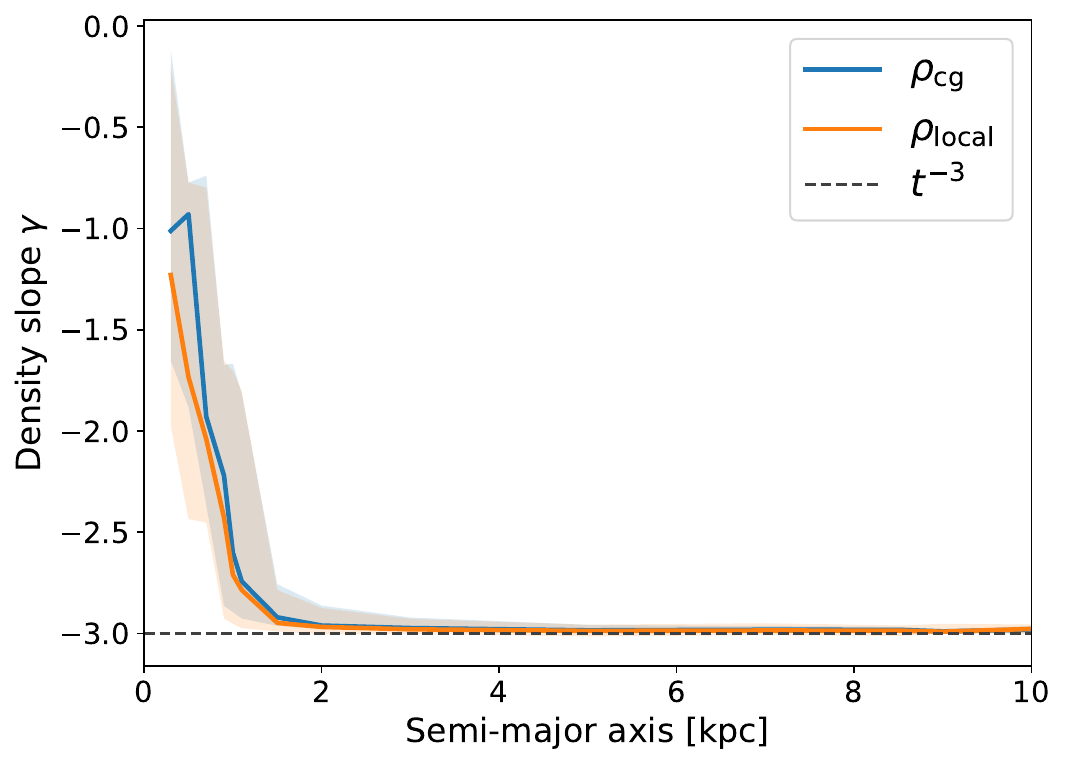}
    \caption{Median logarithmic density slope $\gamma$ as a function of semi-major axis for the coarse-grained density $\rho_{\rm cg}$ and the local-density estimator $\rho_{\rm local}$. The dashed line corresponds to the ballistic scaling $\gamma=-3$. The shaded band corresponds to the $1\sigma$ spread among Monte Carlo realizations.}
    \label{fig:slope_vs_radius}
\end{figure}

The radial dependence of the fitted density slope $\gamma$ from Eq.~\eqref{eq:gamma} is shown in Fig.~\ref{fig:slope_vs_radius}. Within the adopted packet initialization and fitted time interval, both the coarse-grained density $\rho_{\rm cg}$ and the local estimator $\rho_{\rm local}$ approach the full-rank ballistic benchmark $\gamma\simeq-3$. The convergence is nearly complete beyond $a\sim2\,{\rm kpc}$, where the two estimators become almost indistinguishable. Significant deviations appear only in the innermost Galactic region, where repeated perturbations and strong orbital shear slow the effective dilution. The reconstructed evolution is consistent with streams behaving as transient freely expanding debris rather than long-lived self-gravitating structures.

\begin{figure}[htb!]
    \centering
    \includegraphics[width=\linewidth]{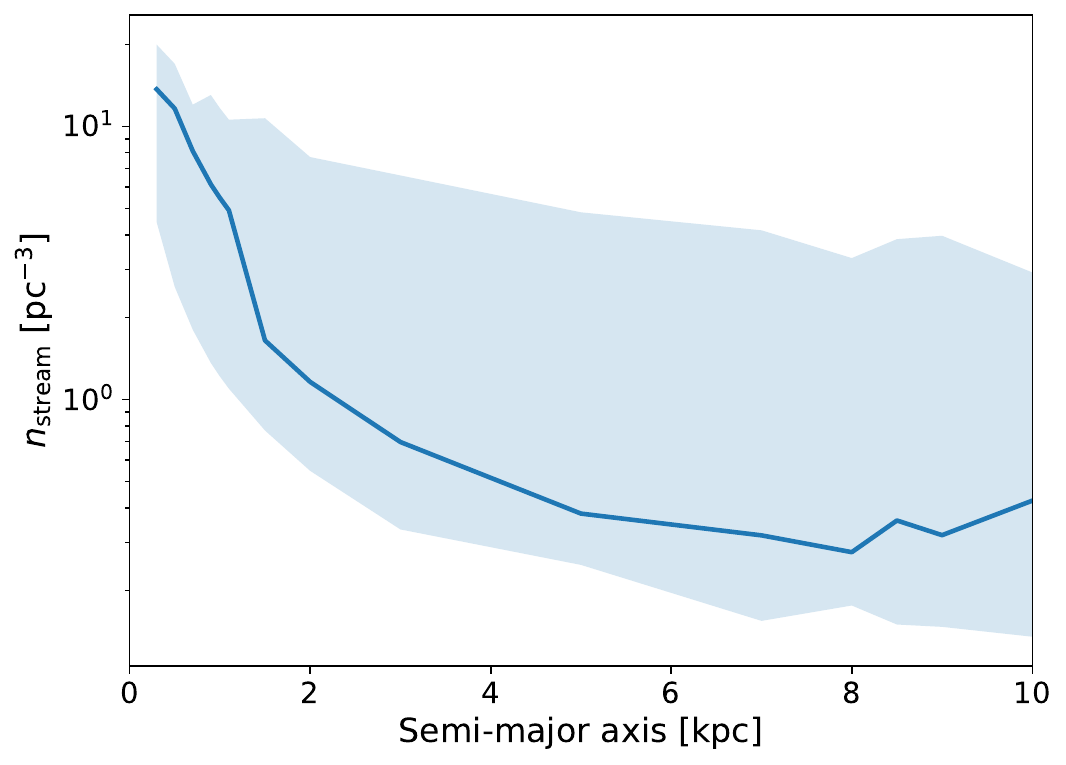}
    \caption{Estimated stream number density as a function of semi-major axis. The shaded region indicates the statistical spread among realizations.}
    \label{fig:nstream_vs_radius}
\end{figure}

The strong dilution of the streams has direct implications for their steady-state abundance. Figure~\ref{fig:nstream_vs_radius} shows the inferred number density of surviving streams, which decreases systematically with galactocentric radius. The broad uncertainty band reflects the sensitivity of the population to rare encounters and stochastic disruption histories.

\section{Discussion}

\begin{figure}[htb!]
    \centering
    \includegraphics[width=\linewidth]{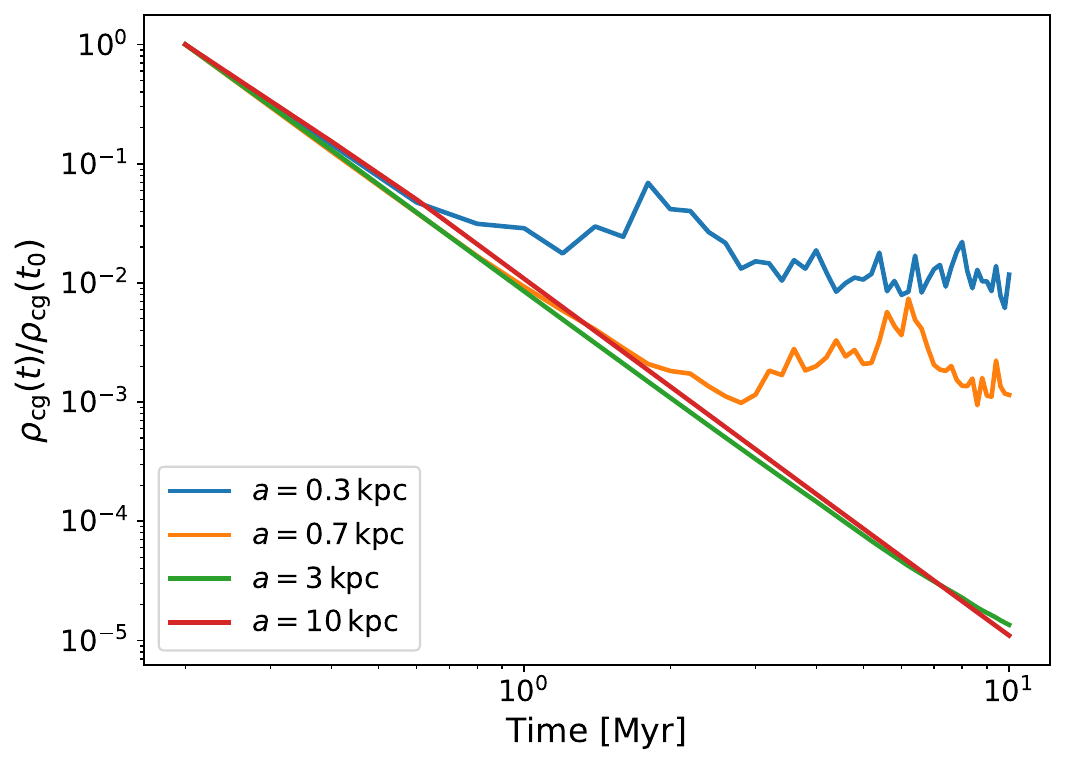}
    \caption{Representative evolution of the normalized coarse-grained tracer-density estimator $\rho_{\rm cg}(t)/\rho_{\rm cg}(t_0)$ for streams formed at $a=0.3\,{\rm kpc}$, $a=0.7\,{\rm kpc}$, $a=3\,{\rm kpc}$, and $a=10\,{\rm kpc}$.}
    \label{fig:density_evolution_compare}
\end{figure}

The reconstructed stream evolution indicates that tidally stripped AMC debris evolves predominantly through collisionless expansion and phase mixing rather than through self-gravitating dynamics. Figure~\ref{fig:density_evolution_compare} shows the normalized evolution of the coarse-grained tracer-density estimator $\rho_{\rm cg}(t)/\rho_{\rm cg}(t_0)$ for representative streams formed at different galactocentric radii. The normalization removes the arbitrary absolute scale associated with the covariance-volume estimator and isolates the relative dilution rate of the reconstructed debris. Outer-halo realizations evolve approximately according to the ballistic scaling $\rho_{\rm cg}\propto t^{-3}$ expected from Eq.~\eqref{eq:rho_ballistic}, while streams formed in the inner Galaxy exhibit stronger temporal fluctuations because of repeated perturbations and orbital shear. The intermediate case at $a=0.7\,{\rm kpc}$ already shows substantial dilution, while still retaining visible fluctuations inherited from the denser stellar environment of the inner halo.

\begin{figure}[htb!]
    \centering
    \includegraphics[width=\linewidth]{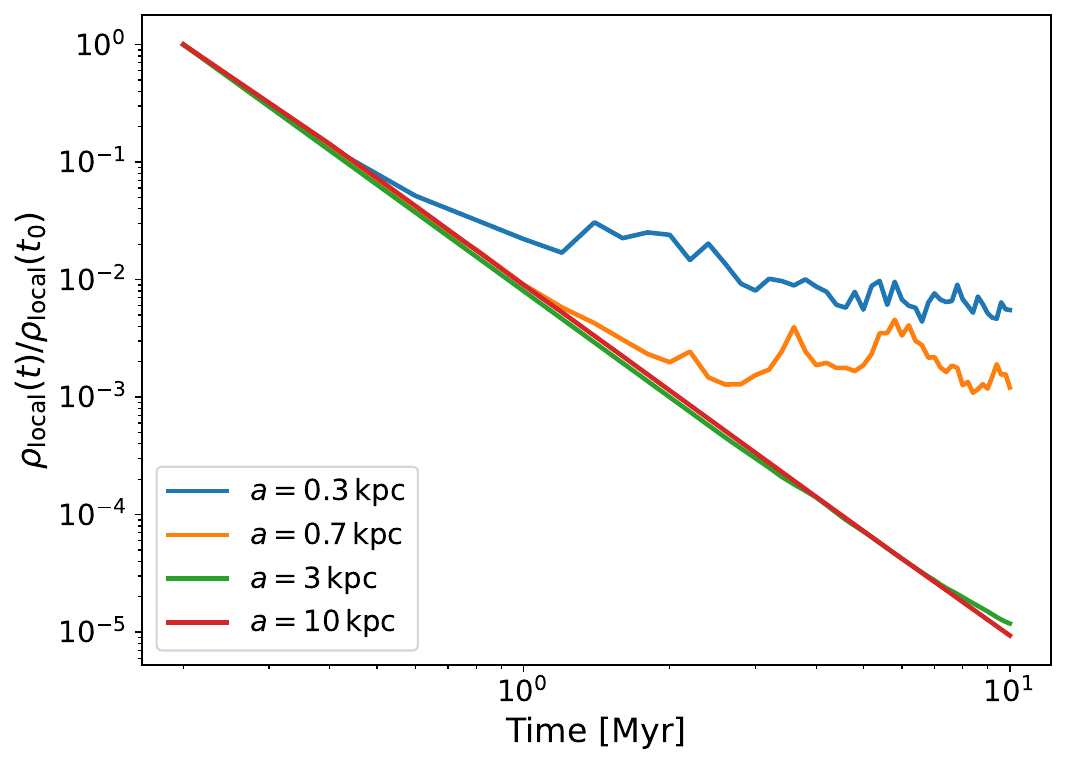}
    \caption{Representative evolution of the normalized local tracer-density estimator $\rho_{\rm local}(t)/\rho_{\rm local}(t_0)$ for streams formed at $a=0.3\,{\rm kpc}$, $a=0.7\,{\rm kpc}$, $a=3\,{\rm kpc}$, and $a=10\,{\rm kpc}$.}
    \label{fig:density_local_evolution_compare}
\end{figure}

The coarse-grained estimator $\rho_{\rm cg}$ should not be interpreted as the microscopic density within individual stream filaments. Instead, it probes the large scale geometric dilution of the tracer ensemble reconstructed from the covariance volume of the stream. To investigate the survival of localized overdensities more directly, Fig.~\ref{fig:density_local_evolution_compare} shows the normalized evolution of the local tracer density estimator $\rho_{\rm local}(t)/\rho_{\rm local}(t_0)$ reconstructed from nearest-neighbor tracer distances. The outer halo realizations again evolve close to the ballistic scaling expected for approximately collisionless free expansion. By contrast, inner halo realizations retain larger local overdensities over longer timescales and develop persistent fluctuations associated with the filamentary morphology of the streams. Localized overdensities can survive substantially longer than suggested by the covariance-volume estimator alone, even while the stream undergoes significant large scale dilution.

The effective steady-state estimate introduced above can be interpreted schematically in terms of a kinetic equation for the stream population,
\begin{equation}
    \frac{{\rm d}n_{\rm stream}}{{\rm d}t} = S - \Gamma_{\rm diffuse}\,n_{\rm stream}\,,
\end{equation}
where $S$ is the local source term associated with AMC disruption and $\Gamma_{\rm diffuse}$ is the effective dilution rate extracted from the reconstructed density evolution. The rapid decline of the reconstructed density estimators indicates that dilution efficiently suppresses the accumulation of a long-lived steady-state stream population throughout most of the Galactic halo.

This framework has direct implications for axion DM haloscope searches~\cite{Sikivie:1983ip,Sikivie:1985yu}. The interpretation of haloscope signals in terms of the underlying Galactic phase-space distribution has been developed extensively in Ref.~\cite{Foster:2017hbq}, including the effects of velocity substructure and time-dependent modulation. In haloscope experiments the signal frequency is determined by the axion kinetic energy, so that the intrinsic rest-frame linewidth scales as
\begin{equation}
    \frac{\Delta\nu}{\nu_a}\sim\frac{\sigma_v^2}{c^2}\,.
\end{equation}
In the detector frame, the observable linewidth is set by the projection of the stream velocity dispersion along the Earth--stream relative motion,
\begin{equation}
    \label{eq:deltanu}
    \Delta\nu_{\rm lab}\simeq \nu_a\,\frac{v_\oplus\,\sigma_{\rm avg}}{c^2}\,,
\end{equation}
where $v_\oplus$ denotes the Earth--stream relative speed, $\sigma_{\rm avg}=\sqrt{(\sigma_l^2+2\sigma_t^2)/3}$, and the axion rest-mass frequency is $\nu_a=m_a/(2\pi)$. Within the present astrophysical population model, rapid density dilution is the main suppression of the expected encounter rate. Experimentally, however, sensitivity to any surviving stream also depends on sufficiently fine spectral resolution and appropriate Doppler tracking~\cite{ADMX:2023ctd, Yi:2025ktq, OHare:2025jpr}.

The comparison with the cavity bandwidth establishes that the signal remains inside the resonator response; it does not determine whether the line is resolved in the recorded spectrum. Spectral resolvability also depends on the Fourier or resolution bandwidth, the coherent integration time, and the treatment of time-dependent Doppler shifts. ADMX has performed a non-virialized search that explicitly accounts for Galactic Doppler shifts and is sensitive to cold flows with velocity dispersion of order $10^{-7}c$~\cite{ADMX:2023ctd, ADMX:2024pxg}. Yi and Ko have demonstrated a multi-resolution analysis spanning the standard-halo, tidal-stream, and big-flow regimes~\cite{Yi:2025ktq}. Studies of fine-grained dark matter structure likewise find that sufficiently high frequency resolution can materially affect haloscope sensitivity~\cite{OHare:2025jpr}.

In the Monte Carlo implementation, a stream is considered visible while its reconstructed density satisfies
\begin{equation}
    \rho_{\rm stream}>\rho_{\rm thr}\,,
\end{equation}
where the threshold is defined relative to the local Galactic DM density $\rho_{\rm loc}=0.45\,{\rm GeV\,cm^{-3}}$ as $\rho_{\rm thr}=10\,\rho_{\rm loc}$. This criterion selects streams that are significantly overdense with respect to the smooth halo background. The corresponding visibility time is the last time at which the stream remains overdense relative to this threshold. The simulations indicate that the retained density fraction can decrease to values as small as $\sim10^{-9}$, while the corresponding visibility timescale is typically
\begin{equation}
    t_{\rm vis}\sim10^5\text{--}10^7\,{\rm yr}\,,
\end{equation}
depending on orbital radius and disruption history. Although streams are continuously produced throughout the Galactic halo, the resulting steady-state population is dominated by diffuse structures with small local density contrasts.

\begin{figure}[htb!]
    \centering
    \includegraphics[width=\linewidth]{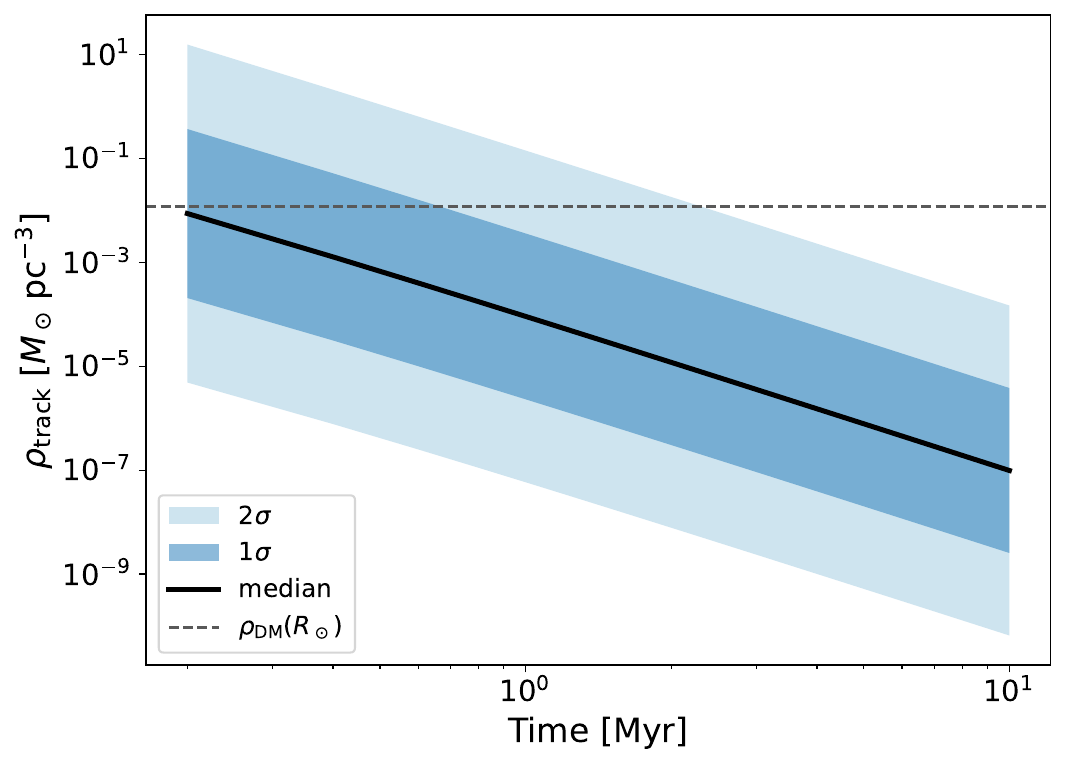}
    \caption{Evolution of the analytic stream-density track $\rho_{\rm track}$ at the Solar radius, $a=8.5\,{\rm kpc}$. The solid black line shows the median over realizations, while the dark and light shaded regions correspond to the $1\sigma$ and $2\sigma$ spreads in log space, respectively. The horizontal dashed line indicates the local Galactic DM density $\rho_\odot$.}
    \label{fig:density_track_evolution_solar}
\end{figure}

To investigate the implications for direct detection more quantitatively, we perform dedicated Solar-neighborhood simulations at $a=8.5\,{\rm kpc}$ with $N_{\rm AMC}=10^6$, including the Earth-relative velocity $v_\oplus=232\,{\rm km\,s^{-1}}$ and an experimental exposure time $T_{\rm exp}=10\,{\rm yr}$. For the conservative prescription in which only fully disrupted AMCs are counted as stream sources, we obtain a disruption fraction $f_{\rm dis}\simeq10^{-4}$ together with characteristic stream masses
\begin{equation}
    \langle M_{\rm stream}\rangle\sim5\times10^{-17}\,M_\odot\,.
\end{equation}
The inferred stream abundance near the Solar circle remains small, $n_{\rm stream}\lesssim{\cal O}(0.1\text{--}1)\,{\rm pc^{-3}}$, with a highly skewed distribution dominated by rare disruption events. The operational stream density entering the population-level visibility estimate is the analytic ballistic density track $\rho_{\rm track}$ introduced in Eq.~\eqref{eq:rho_ballistic}. Figure~\ref{fig:density_track_evolution_solar} shows its evolution at the Solar radius. The solid curve corresponds to the median over realizations, while the dark and light shaded regions indicate the $1\sigma$ and $2\sigma$ spreads in log space, respectively. The comparison with the local Galactic DM density $\rho_\odot$ illustrates the rapid dilution of the stream population relative to the smooth halo background. While the median stream density rapidly falls below $\rho_\odot$, the broader $2\sigma$ region shows that a subset of realizations can temporarily retain substantially larger densities because of rare disruption histories.

The resulting encounter-probability distribution is similarly sparse. In the conservative prescription, the median encounter probability over a decade-long exposure is effectively zero. At the Solar radius, only a fraction of order $10^{-4}$ of the realizations yields a nonzero encounter probability, with a nonzero-realization median $P_{\rm enc}\sim10^{-11}$ and an upper tail reaching $P_{\rm enc}\sim10^{-6}$. Observable events are associated with rare realizations involving comparatively young and nearby streams rather than with a persistent background of long-lived overdense substructure. This picture is qualitatively consistent with the broader post-inflationary scenario in which direct-detection experiments probe a highly intermittent axion density field containing both rare overdense structures and extended underdense minivoids~\cite{Eggemeier:2022hqa}.

\begin{figure}[htb!]
    \centering
    \includegraphics[width=\linewidth]{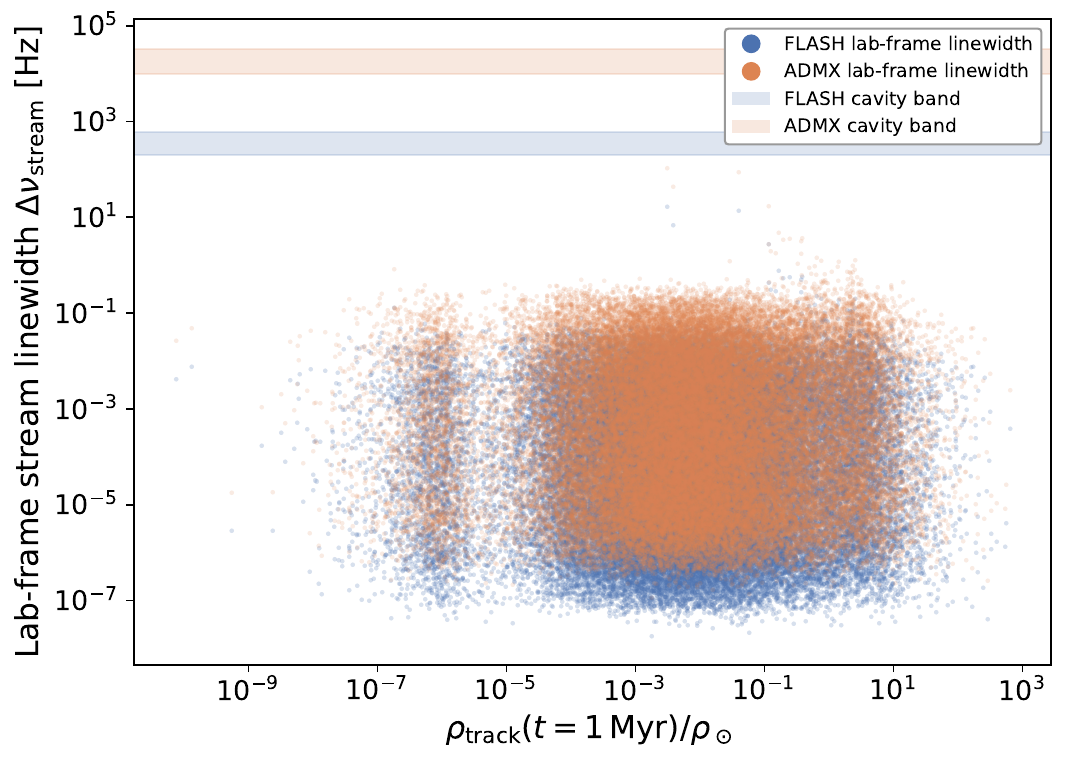}
    \caption{Detector-frame stream linewidth as a function of the reconstructed density contrast at the Solar radius, $a=8.5\,{\rm kpc}$, evaluated at $t=1\,{\rm Myr}$. The shaded horizontal bands indicate the cavity bandwidth ranges for FLASH (blue) and ADMX (orange).}
    \label{fig:haloscope_linewidth_density}
\end{figure}

The corresponding haloscope phenomenology is illustrated in Fig.~\ref{fig:haloscope_linewidth_density}, which shows the detector-frame stream linewidth as a function of the reconstructed density contrast at the Solar radius. The linewidths estimated from Eq.~\eqref{eq:deltanu} are compared with the cavity bandwidths of representative haloscope configurations, namely the FINUDA magnet for Light Axion SearcH (FLASH)~\cite{Alesini:2023qed}, with frequency range $f=100$--$300\,{\rm MHz}$ and quality factor $Q_{\rm cavity}=5\times10^5$, and the Axion Dark Matter eXperiment (ADMX)~\cite{ADMX:2001dbg}, in the configuration probing $f=0.6$--$2\,{\rm GHz}$ and with $Q_{\rm cavity}=6\times10^4$~\cite{ADMX:2018gho, ADMX:2021mio, ADMX:2024pxg}. Although the detector-frame linewidths are substantially larger than the intrinsic rest-frame estimates, they remain many orders of magnitude smaller than the corresponding cavity bandwidths. The stream signal appears effectively monochromatic at the level of the cavity response.

Three distinct spectral scales enter the problem: the intrinsic stream linewidth determined by the velocity dispersion, the secular Doppler drift induced by the Earth--stream relative acceleration, and the instrumental cavity bandwidth. The secular drift of the signal frequency may be estimated as
\begin{equation}
    \dot{\nu}\sim \nu_a\frac{v_\oplus a_\oplus}{c^2}\,,
\end{equation}
where $a_\oplus$ denotes the characteristic acceleration of the Earth in the Galactic frame. This defines an effective drift quality factor,
\begin{equation}
    \Delta\nu_{\rm drift}(T_{\rm int}) \simeq |\dot{\nu}|T_{\rm int}, \qquad
    Q_{\rm drift}(T_{\rm int}) \equiv \frac{\nu_a}{\Delta\nu_{\rm drift}(T_{\rm int})}\,,
\end{equation}
showing that the secular Doppler drift remains far smaller than the instrumental response width of present haloscope cavities. The drift does not significantly broaden the signal at the cavity-response level, although it may become relevant for ultra-long coherent integrations, narrow Fourier-bin analyses, and phase-sensitive signal reconstruction.

The linewidth estimates above also determine the characteristic coherence time of the signal,
\begin{equation}
    t_{\rm coh}\sim\frac{1}{\Delta\nu_{\rm lab}}\,.
\end{equation}
For the stream realizations reconstructed here, the narrow velocity dispersions imply coherence times ranging from fractions of a second up to $\sim10^7\,{\rm s}$ for the narrowest streams. In realistic searches, however, the effective coherence time relevant for matched filtering and phase-sensitive analyses is also affected by finite detector resolution and by the time-dependent Doppler modulation induced by the Earth--stream relative motion.

The approximately rectangular support of the Monte Carlo distribution in Fig.~\ref{fig:haloscope_linewidth_density} reflects the weak correlation between the density contrast and the velocity width in the present framework. The linewidth is controlled primarily by the internal velocity dispersion projected along the Earth--stream relative motion, while the density contrast depends mainly on the stripped mass fraction and on the subsequent geometric dilution during ballistic expansion. Density dilution and spectral narrowness are largely decoupled observables, so even strongly diluted streams can remain spectrally narrow in the laboratory frame. The bounded support of the distribution is also influenced by the finite AMC mass range and by the fixed Solar-radius benchmark adopted in the present analysis.

The direct-detection phenomenology proposed in Ref.~\cite{OHare:2023rtm} remains viable, but is restricted to the rare high-density tail of the stream population. In this regime the signal would appear as a transient and extremely narrow spectral feature with slowly varying Doppler modulation induced by the Earth--stream relative motion. The intrinsic evolution of the stream morphology occurs on Galactic dynamical timescales vastly longer than experimental integration windows, allowing the signal to be treated as effectively stationary during a typical observation.

Several limitations of the present analysis should nevertheless be emphasized. The Monte Carlo framework treats the stream evolution in an effective manner and neglects possible subsequent perturbations from Galactic substructure, giant molecular clouds, and time-dependent non-axisymmetric structures. In addition, the present treatment follows the debris in the collisionless approximation and does not attempt to resolve interference effects or fine-grained caustic structures associated with the wave nature of the axion field. The predicted linewidths should be interpreted as idealized collisionless estimates, since unresolved perturbations, wave interference effects, finite detector resolution, and experimental systematics may broaden the observable signal under realistic observational conditions.

Future work should extend the present framework in several directions. Higher-resolution tracer reconstructions and dedicated $N$-body simulations would allow a more precise characterization of the long-term stream morphology and dilution rates, while more realistic Galactic potentials could quantify the impact of disk crossings and time-dependent perturbations. A particularly important extension would be to incorporate the wave nature of the axion field directly into the stream evolution. Interference effects, coherent wave dynamics, and fine-grained caustic structures could modify the local phase-space distribution on scales inaccessible to the present collisionless treatment. It would also be interesting to couple the present framework directly to mock haloscope pipelines in order to quantify the detectability of transient stream signals in realistic experimental searches.

\section{Conclusions}

In this work, we have investigated the formation and evolution of axion streams generated by the tidal disruption of axion miniclusters through stellar encounters in the Galactic halo. Combining a large-scale Monte Carlo treatment of cumulative stellar flybys with a tracer reconstruction of the stripped debris, we have followed the subsequent evolution of the streams across a broad range of galactocentric radii and characterized their masses, morphology, kinematics, and density evolution. Our analysis identifies a simple dynamical picture for the fate of tidally stripped AMC debris. Although stellar encounters can efficiently remove material from miniclusters, the stripped component typically carries only a modest fraction of the progenitor mass while inheriting velocity scales associated with the deeper potential of the original bound configuration. The characteristic kinetic energy of the debris generally exceeds its residual self-gravitational binding energy, so that the subsequent evolution is governed primarily by collisionless expansion and orbital shear rather than by self-gravitating dynamics.

The simulations reveal a transition between two dynamical regimes. In the inner Galaxy, repeated stellar encounters continuously perturb the streams, producing strongly filamentary structures with large axis ratios and substantial temporal fluctuations in density and morphology. By contrast, streams formed at larger galactocentric radii evolve more smoothly and rapidly approach an approximately ballistic regime. In this limit, both the local and coarse-grained density estimators approach the full-rank ballistic benchmark over the simulated interval expected from free phase-space dilution, while the surviving density fraction decreases to extremely small values at late times. The resulting steady-state abundance of dense streams near the Solar circle is strongly suppressed.

These results have direct implications for axion haloscope searches. The reconstructed stream velocity dispersions imply extremely narrow spectral features compared with both the signal expected from the virialized Galactic halo and the cavity response bandwidths of current experiments. For the FLASH and ADMX benchmark configurations considered here, the detector-frame stream linewidths are many orders of magnitude smaller than the corresponding cavity bandwidths and therefore occupy only a small fraction of the resonator response. Within the present population model, the main astrophysical suppression is the small probability of encountering a sufficiently dense stream during the lifetime of an experiment. Detecting and spectrally resolving such a signal nevertheless requires sufficiently fine frequency resolution, an appropriate coherent-integration strategy, and explicit treatment of its time-dependent Doppler modulation. Our dedicated Solar-neighborhood realizations indicate that potentially observable events are controlled by the rare high-density tail associated with comparatively recent and nearby disruption events, rather than by a quasi-steady population of persistent overdense streams.

More broadly, our results show that the phenomenology of axion substructure is intrinsically dynamical. The observational relevance of streams depends not only on the formation of compact miniclusters, but also on the subsequent competition between tidal stripping, orbital shear, and phase-space dilution. Within the collisionless framework developed here, the stripped debris evolves predominantly into diffuse filamentary structures whose detectability is strongly suppressed by rapid density dilution despite their extremely cold kinematics.

Several important extensions nevertheless remain to be explored. Higher-resolution $N$-body simulations and more realistic Galactic potentials could provide a more precise characterization of the long-term stream evolution and dilution rates. A particularly important direction will be to incorporate the wave nature of the axion field directly into the stream dynamics, including possible interference effects and fine-grained caustic structures inaccessible to the present collisionless treatment.

Overall, our results provide a unified dynamical picture of the interplay between tidal stripping and stream evolution in post-inflationary axion cosmology. While AMCs may survive as compact substructures throughout the Galactic halo, the debris generated by their disruption evolves predominantly into transient diffuse streams whose observational relevance is controlled by rare high-density realizations. Future high-resolution simulations and dedicated experimental searches will determine whether such transient axion streams can ultimately be observed.

\vspace{.3cm}

\section*{DATA AVAILABILITY}

The source code and analysis tools used to generate the results presented in this article are publicly available in Ref.~\cite{visinelli_2026_22756132}.
\vspace{.3cm}

\begin{acknowledgments}
We thank Ciaran O'Hare and Giovanni Pierobon for stimulating conversations. LV acknowledges support by the National Natural Science Foundation of China (NSFC) through the grant No.~12350610240 ``Astrophysical Axion Laboratories''. This work is supported by the Istituto Nazionale di Fisica Nucleare (INFN) through the Commissione Scientifica Nazionale~4 (CSN4) Iniziativa Specifica ``Quantum Universe'' (QGSKY). This publication is based upon work from the COST Actions ``COSMIC WISPers'' (CA21106) and ``Addressing observational tensions in cosmology with systematics and fundamental physics (CosmoVerse)'' (CA21136), both supported by COST (European Cooperation in Science and Technology). MN acknowledges the support of BNSF KP-06-N98/2 from 01.12.2025 ``Nuclear spectra, isomers and symmetry: test of fundamental constants and dark matter'' and BNSF/SU 80-10-180 from 20.03.2026 ``Experimental detection of dark matter''. 
\end{acknowledgments}

\bibliographystyle{apsrev4-1}
\bibliography{references}

\end{document}